\documentclass[12pt]{iopart}
\usepackage{graphicx}
\usepackage{iopams}

\newcommand{\vf}{\bar v_\mathrm{f}}

\begin{document}

\title{Spreading of a density front in the K\"{u}ntz-Lavall\'{e}e model of porous media}

\author{Maciej Matyka and Zbigniew Koza}

\address{Institute of Theoretical Physics, University of Wroc{\l}aw,
pl.\ Maksa Borna 9, 50-204 Wroc{\l}aw, Poland,}
\ead{maq@ift.uni.wroc.pl}
\begin{abstract}
We analyze spreading of a density front in the K\"{u}ntz-Lavall\'{e}e model of
porous media. In contrast to previous studies, where unusual properties of the
front were attributed to anomalous diffusion, we find that the front evolution
is controlled by normal diffusion and hydrodynamic flow, the latter being
responsible for apparent enhancement of the front propagation speed. Our
finding suggests that results of several recent experiments on porous media,
where anomalous diffusion was reported based on the density front propagation
analysis, should be reconsidered to verify the role of a fluid flow.

%spreading process in systems with concentration dependent diffusion coefficient
%using Lattice Gas Automata (LGA) model. We isolate and subtract hydrodynamical
%flow effect from overall transport mechanism. We show that in a moving frame of
%reference the density front spreading shows typical Fickian behavior and scales
%with $1/2$ time exponent.
\end{abstract}

\submitto{\JPD}
%\maketitle

\section{Introduction}

The root-mean-square displacement, $\sqrt{\langle r^2\rangle}$, of a
single diffusing particle at long times $t$ is typically proportional
to $t^{\alpha}$ with $\alpha = 1/2$. The same scaling law is also
usually obeyed by the width of a region infiltrated by a diffusing
front. However, there are also many natural phenomena where the
exponent $\alpha$ differs from $1/2$. These include, for example,
water absorption in porous building materials \cite{ghandy,pel,kuntz-JPD01,Lockington},
copper sulfate diffusion into deionized water
\cite{Carey95,kuntz-rapidcomm}, transport in high polymer
systems \cite{Rehage70}, diffusion through synthetic membranes
\cite{Fowlkes}, and some aspects of surface diffusion \cite{Naumovec}.
Diffusion processes with $\alpha = 1/2$ are ubiquitous in Nature and
hence called `normal', whereas those with $\alpha \neq 1/2$ are quite
rare and called `anomalous'.

There are several mechanisms that are known to bring about anomalous
tracer \cite{Gomer} diffusion; for example, a power-law probability
distribution of waiting times between individual steps of a random
walk or a power-law probability distribution of distances covered by
the diffusing particle at each jump
\cite{HavlinBook,Metzler-Klafter}. However, far less is known about
anomalous chemical \cite{Gomer} diffusion, i.e.\ diffusion of
macroscopic quantities of matter. Anomalous spreading of diffusion
fronts is a many-body effect that results from complicated
interactions between diffusing particles as well as between the
particles and the surrounding (e.g. porous or fractal-like) medium.
In many cases the nature of these interactions remains unknown, so
actually there is no satisfactory explanation of the anomalous
diffusion in these systems \cite{Naumovec}. Thus, any progress in the
theory of anomalous chemical diffusion would be much welcomed.

Recently K\"{u}ntz and Lavall\'{e}e
\cite{kuntz-rapidcomm,kuntzJPD03,kuntz-PRE05} suggested that
anomalous chemical diffusion can be a common phenomenon in systems
with concentration-dependent diffusion coefficient of the diffusing
entity. They based it on extensive computer simulations of a
lattice-gas automata (LGA) model \cite{kuntzJPD03,kuntz-PRE05}, and a
refined analysis of a CuSO$_4$ front diffusion into deionized water
experiment \cite{kuntz-rapidcomm}. Were this conjecture correct, it
could be used to identify the cause of anomalous spreading of
concentration fronts in many real systems. %, like ****.

In this paper we focus on the K\"{u}ntz and Lavall\'{e}e (KL) model
and propose a much simpler interpretation of the simulation results
obtained in \cite{kuntzJPD03,kuntz-PRE05}. Our approach  indicates
that the diffusion in the KL model is normal. The key to the proper
interpretation of the simulation results is the fact that the KL
model is an extension of the Frish-Hasslacher-Pomeau (FHP) lattice
gas model of fluid flow. Even though K\"{u}ntz and Lavall\'{e}e
applied it in a porous medium, which significantly reduced the
 flow velocity, hydrodynamic effects were not eliminated altogether.
 We found that the fluid flow velocity is large enough to enhance the
effective speed of the diffusing front, apparently enlarging the value of the
scaling exponent $\alpha$ above $1/2$. We show that when the flow is taken into
account, the exponent $\alpha$ assumes its expected, ``normal'' value $1/2$.

%We analyze the density front spreading process in non saturated porous media.
%Diffusive processes in systems with concentration dependent diffusion
%coefficient has been previously investigated both experimentally \cite{} and
%theoretically \cite{}. We show that in a moving frame of reference density
%front spreading is a Fickian process which scales with $1/2$ exponent.

In a broader perspective, our study shows that  application of ``hydrodynamic''
lattice-gas automata models to study diffusion in porous media requires a great
caution. Special attention must be paid to modelling the porosity of the medium
to ensure that there is no macroscopic flow through the system. If this is
impossible, the flow, even of minute magnitude, must be explicitly taken into
account in the analysis of results.

The structure of the paper is as follows. In Section~\ref{sec:model} we briefly
introduce the  K\"{u}ntz-Lavall\'{e}e  model and the main results obtained thus
far. In Section~\ref{sec:results} we present our interpretation of these
results, and show that actually there is no anomalous diffusion in the system.
Section \ref{sec:discussion} presents discussion of the major properties of the
model.  Finally, Section \ref{sec:conclusions} is devoted to conclusions.

\section{The K\"{u}ntz-Lavall\'{e}e model
\label{sec:model}}

\begin{figure}[!htb]
\centering
  \includegraphics[scale=0.7]{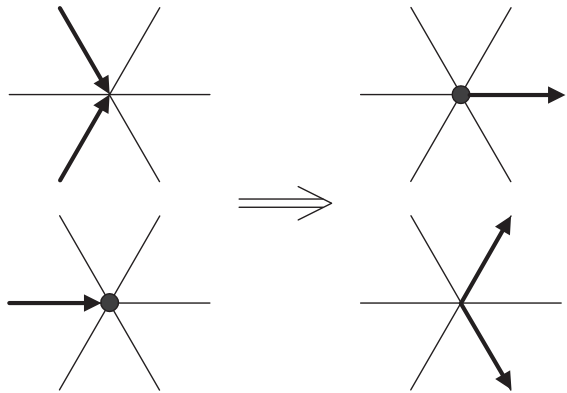}
  \caption{Collision rules in the FHP$_5$ model.
           Filled circles represent particles at rest.
  \label{pic:collisionrules}}
\end{figure}

The K\"{u}ntz-Lavall\'{e}e model
\cite{kuntz-rapidcomm,kuntzJPD03,kuntz-PRE05,kuntzTPM01} describes
diffusion of a fluid in a porous medium and is an extension of a
deterministic Frish-Hasslacher-Pomeau (FHP$_5$) lattice-gas automata model
of fluid flow \cite{frish-lga}. The system is reduced
to a two-dimensional triangular lattice, $L_x$ lattice units (l.u.)
long and $L_y\sqrt{3}/2$ l.u.\ wide. The total number of lattice
nodes is thus $N = L_x L_y$. They are occupied by indistinguishable
particles, whose velocities belong to a discrete, 7-element set
(Fig.~\ref{pic:collisionrules}). Any particles occupying the same
lattice site must have different velocities, so that each node can be
occupied by at most 7 particles. The time is discrete and at each
time step particles either stay at rest (if their speed is 0) or jump
to the adjacent site pointed at by their velocity vector. Particles
arriving simultaneously at the same node may collide. The collision
rules for the FHP$_5$ model are depicted in
Fig.~\ref{pic:collisionrules}; they conserve the mass and momentum.
Additionally, a porous medium is modelled by assuming that a fraction
$c_s$ of randomly chosen lattice nodes is permanently occupied by
the so called specular scatterers (such scatterers
behave like tiny rigid squares parallel to the ``x'' axis \cite{kuntzTPM01}).
The system is confined between rigid, impenetrable walls along the
``x'' direction, whereas periodic boundary conditions are assumed
along the ``y'' direction. A characteristic feature of the model thus
constructed is a strong dependence of the diffusion coefficient $D$
on the reduced local particle concentration $c$ \cite{kuntzJPD03},
which is defined as the average number of particles per node divided
by 7. It turns out that $D$ assumes a minimum value at $c\approx 0.2$
and diverges to infinity as $c$ goes either to $0$ or $1$
\cite{kuntzJPD03}.

Initially the system is uniformly filled with particles so that the
reduced concentration is $c_2$ throughout the system. To this end
$7c_2N$ particles are distributed randomly in the system in such a
way that their mean velocity vanishes and any particles occupying the
same node have different velocities. Next, a narrow strip of nodes at
the left-hand side of the system is uniformly filled with additional
particles so as to increase the reduced concentration to $c_1 > c_2$. In
this way a step-like nonuniform initial condition is formed,
\begin{equation}
  \label{eq:ini-cond}
    c(x,0) = c_1H(-x), \;x<0, \quad c(x,0) = c_2 H(x), \;x> 0,
\end{equation}
where $H(x)$ is the Heviside step function. The origin of the reference system
is set to the interface between the areas of high and low initial
concentration, so that $x=0$ corresponds to the initial location of the step.

During the simulation, by injecting new particles if necessary, the reduced
concentration in the boundary strip $x<0$ is maintained fixed at $c_1$. This
corresponds to boundary conditions that are often used in studies on water
infiltration into a porous medium:
\begin{equation}
\label{eq:boundary-cond}
 c(x,t) = c_1\; \mathrm{for}\; x=0, \qquad c(x,t) \to c_2
\;\mathrm{as}\; x \to \infty.
\end{equation}
%Note that if the diffusion coefficient $D$ was constant, the solution to the
%diffusion equation with these boundary conditions would read
%\begin{equation}
%   c(x, t) = c_2 + (c_1-c_2)\,\mbox{erf}\,(x/\sqrt{4Dt}), \quad  (x \ge 0),
%\end{equation}
%where $\,\mbox{erf}\,(z) = 2\pi^{-1/2}\int_0^z \exp(-\xi^2)\,\mathrm{d}\xi$ is
%the error function. This formula describes a mobile concentration front whose
%form obeys, for all times $t$, the normal diffusion scaling $c(x,t) =
%g(x/t^\alpha)$ with $\alpha = 1/2$ and some scaling function $g$. Our aim is to
%verify if the concentration front obeys the same scaling in the above-defined
%LGA model.
%
%

\section{Results}
\label{sec:results}

\begin{figure}
  \centering
    \includegraphics[scale=0.5]{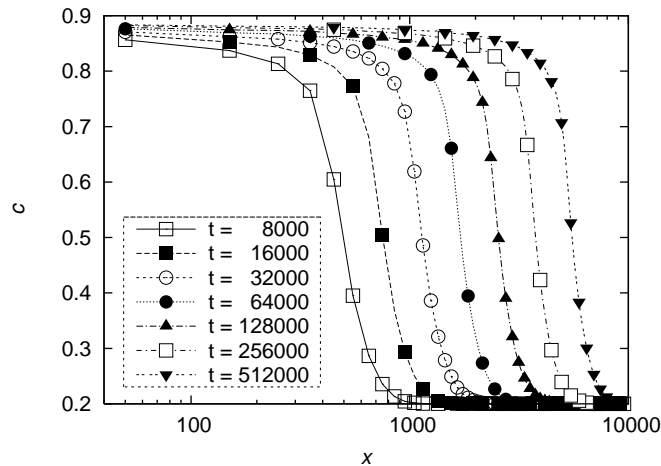}
    \caption{Reduced concentration profile $c(x,t)$ as a function of position $x$
             for $t = 8000 \times 2^k$ time steps, $k = 0,1,\ldots,6$.
             The boundary conditions are $c_1 = 0.9$ and $c_2 = 0.2$.
  \label{fig:densityprofiles}}
\end{figure}

We performed extensive simulations of a concentration front
propagation using a triangular lattice with $L_x=8000$, $L_y = 200$,
the density of scatterers $c_s = 0.08$, and the initial and boundary
conditions determined by $c_1=0.9$ and $c_2=0.2$. The results for $t
= 8000 \times 2^k$, $k = 0,1,\ldots,6$, are depicted in Figure
\ref{fig:densityprofiles} as a semi-logarithmic plot. Both the shape
of the consecutive curves and the distances between them appear to be
almost the same, which suggests that the concentration profile
$c(x,t)$ can actually be expressed as a function  of a single
variable $x/t^\alpha$.

Validity of the scaling hypothesis can be verified by plotting several
concentration profiles, obtained at different times, as functions of
$x/t^\alpha$. If the scaling holds, all these plots should collapse into a
single curve.
Such analysis was already performed in \cite{kuntzJPD03}, where it was found
that $\alpha \approx 0.55$. This was interpreted as an evidence that the
transport in the LGA model is superdiffusive ($\alpha > 1/2$).

This finding, however, leads to a paradox: how addition of scatterers can
\emph{enhance} transport of particles? According to K\"{u}ntz and Lavall\'{e}e
\cite{kuntzJPD03}, this paradox can be explained by two hypotheses. The first
one attributes superdiffusion of particles to a very strong dependence of their
diffusivity on concentration. The second one says that anomalous diffusion is
actually a transient phenomenon: in the limit of time $t\to\infty$ the scaling
exponent $\alpha$ will \emph{extremely slowly} decrease to $1/2$.

However, this argumentation neglects a well known fact, widely used in the
Boltzmann-Matano method  \cite{Gomer,CrankBook}, that any solution to the
diffusion equation
\begin{equation}
 \label{eq:diffusion-equation}
  \frac{\partial c(x,t)}{\partial t}
     =
  \frac{\partial}{\partial x}
     \left(
       D(c)\frac{\partial c(x,t)}{\partial x}
     \right),
\end{equation}
with the initial and boundary conditions (\ref{eq:ini-cond}) and
(\ref{eq:boundary-cond}), satisfy the $x/t^{1/2}$ scaling for all $t$,
irrespective of the form of $D(c)$ \cite{CrankBook}. Consequently, absence of
this scaling indicates that equation (\ref{eq:diffusion-equation}) cannot give
a proper description of the concentration front dynamics in the model. Besides
diffusion, there must be another transport mechanism that controls particle
infiltration into the low-concentration area. We conjecture that this is the
fluid flow induced by a pressure gradient between the high- and
low-concentration areas of the system.

\begin{figure}
 \centering
 \includegraphics[height=0.45\textwidth]{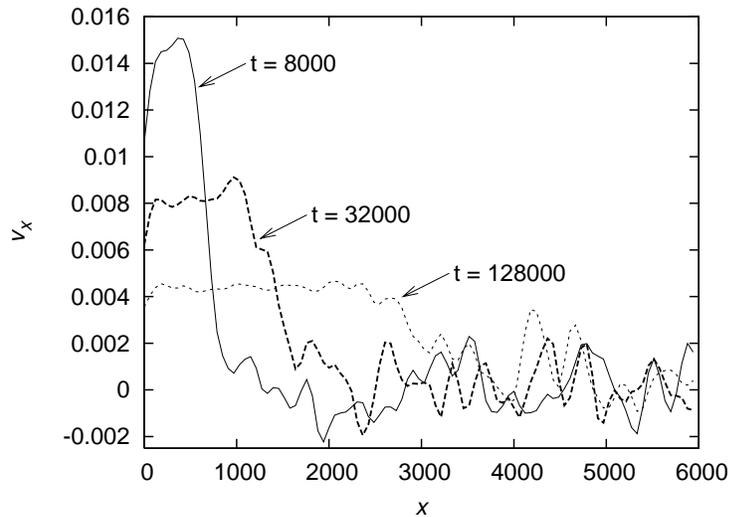}
   \caption{The ``x'' component of the velocity, $v_x$, as a function of the
            distance from the origin, $x$, for time
            $t = 8000\times4^k, k = 0,1,2$.
 \label{fig:velocity-profiles}}
\end{figure}

To check this hypothesis, we started from detailed analysis of spacial and
temporal properties of the ``x'' component of the velocity vector, $v_x(x,t)$.
Figure \ref{fig:velocity-profiles} depicts the velocity profile at several
times $t$. It shows that the region $x>0$ can be split into two parts
characterized by different behavior of $v_x$. The first region coincides with
the particle infiltration area. The system already ``feels'' the pressure
gradient there and responds to it by developing a small but definitely
nonvanishing flow towards the low concentration area. The second region is
located further away from the infiltration area. The system remains there close
to the initial state, with $v_x$ fluctuating around $0$.

In order to study the concentration front dynamics, one thus have to take into
account the front velocity, $v_\mathrm{f}$. In general, this quantity is both
$x$- and $t$-dependent. However, we are interested only in its approximate
value at time $t$, and estimate it as a mean weighted with $c(x,t) - c_2$,
\begin{equation}
 \label{eq:def:v-mean}
  \vf(t) =
     \frac{\displaystyle \int_0^\infty v_x(x,t)[c(x,t) - c_2]\,\mathrm{d}x}%
          {\displaystyle \int_0^\infty [c(x,t) - c_2]\,\mathrm{d}x}.
\end{equation}
In this definition we utilize the fact that
the weight function quickly vanishes outside the front region
(Fig.~\ref{fig:densityprofiles}), while the averaged quantity, $v_x$,
remains almost constant there (Fig.~\ref{fig:velocity-profiles}).
Moreover, the integral form of the definition helps to significantly  reduce the
influence of relatively large statistical noise present in the data for $v_x$.

\begin{figure}
  \centering
  \includegraphics[height=0.45\textwidth]{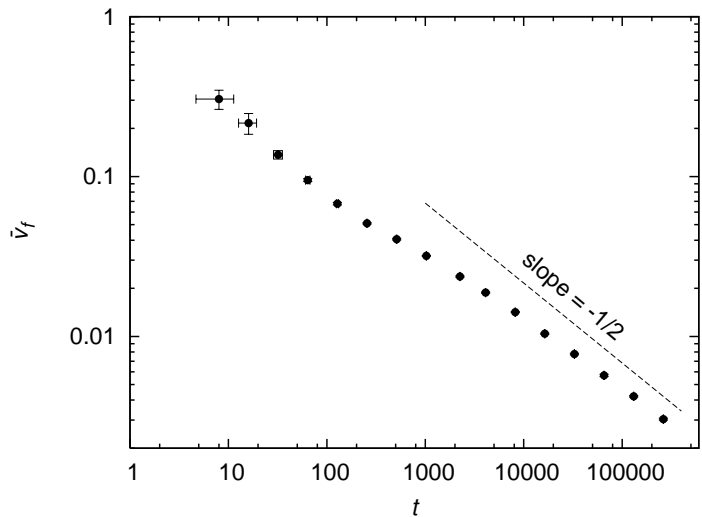}
  \caption{The mean front velocity, $\vf$, as a function of time.
             The dashed line is a guide to the eye with a slope $-1/2$.
    \label{fig:v-t} }
\end{figure}

Figure \ref{fig:v-t} shows that the initial magnitude of $\vf$ is very high,
almost reaching the maximum possible value $1$. At larger times $\vf$ decreases
as $t^{-\beta}$ with $\beta \lesssim 1/2$ very weakly depending on $t$.
Consequently, the distance the front moves due to the flow,
\begin{equation}
 \label{eq:def-s}
   s(t) = \int_0^t \vf(\tau)\,\mathrm{d}\tau,
\end{equation}
for large $t$ grows as $t^{1-\beta(t)}$, i.e.\ a bit faster than the typical
diffusive length $ \approx t^{1/2}$. This suggests that the front dynamics is
controlled by \emph{both} diffusion and flow. Actually, our simulations showed
that after $t=256000$ time steps the front displacement was $\Delta x\approx
3750$ l.u.\ (see Figure \ref{fig:densityprofiles}), of which $s(t) \approx
1440$ l.u, that is, about 38\%,  was caused by the fluid flow.

\begin{figure}
  \centering
  \includegraphics[ width=0.6\textwidth]{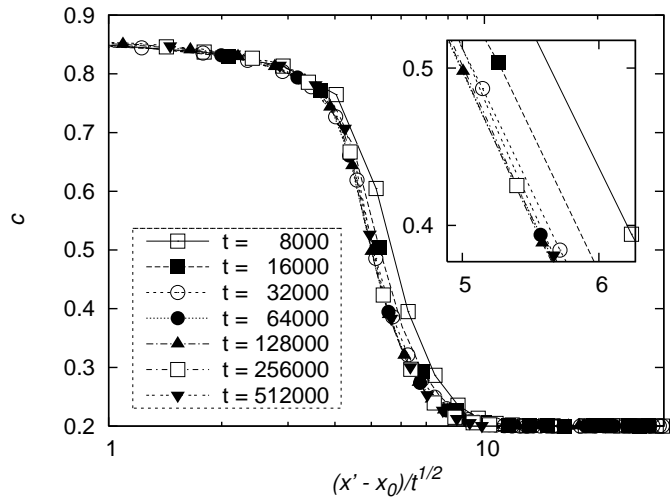}
  \caption{Reduced concentration $c$ as a function of $(x'-x_0)/\sqrt{t}$,
           with $x_0 = -190$. The inset shows an enlargement of the central part of the
           main plot, with linear axes.
     \label{fig:final-scaling} }
\end{figure}

To separate diffusion from flow, we investigated the front propagation using a
mobile reference frame $(x', t)$, with
\begin{equation}
 \label{eq:def-xprime}
  x' \equiv x - s(t).
\end{equation}
This new coordinate system moves along the ``x'' axis with velocity $\vf(t)$,
so as to minimize effects of the fluid flow. Assuming that in this new
reference system the front dynamics is dominated by (normal) diffusion, the
concentration profiles should asymptotically obey a similarity relation
\begin{equation}
  \label{eq:scaling-moving}
  c(x',t) = f((x' - x_0)/\sqrt{t}),
\end{equation} where $f$ is a similarity function and $x_0$ is a
constant. To verify this conjecture, we first used the data for
$c(x',t) = 0.5$ to estimate $x_0\approx -190$, and then plotted $c$
as a function of $(x'-x_0)/\sqrt{t}$ for several times $t$. The
results, obtained for $8000\le t\le512000$, are shown in Figure
\ref{fig:final-scaling}. They confirm that the concentration profiles
asymptotically satisfy~(\ref{eq:scaling-moving}), as for $t\ge64000$
the scaled-up profiles are practically indistinguishable.

\begin{figure}
  \centering
  \includegraphics[width=0.65\textwidth]{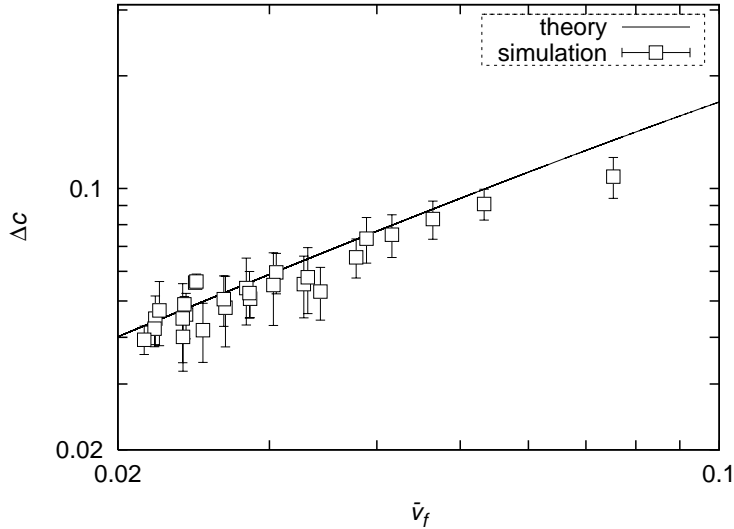}
  \caption{Reduced concentration drop, $\Delta c$, at the boundary $x=0$ as a
           function of the front velocity $\vf$, obtained from simulations
           ($\square$) and equation (\protect\ref{eq:Dc-Dv}) (solid line).
  \label{fig:Dc-Dv} }
\end{figure}

Hydrodynamical flow can also explain another feature of the
model---discontinuity of fluid concentration at the boundary $x = 0$. Even
though the boundary condition~(\ref{eq:boundary-cond}) forces the concentration
to be fixed at $c_1$ for $x < 0$, its values at $x>0$ are significantly smaller
than $c_1$ (see Figure (\ref{fig:densityprofiles}), obtained for $c_1 = 0.9$).
Discontinuity in $c$, $\Delta c$, is related to discontinuity in the flow
velocity $\Delta v \approx \vf$ ($v_x = 0 $ for $x<0$ and $v_x \approx \vf$ for
$x \gtrsim 0$). Let $c^+ = c_1 - \Delta c$ denote the concentration at $x = 1$
l.u. The rate of diffusive particle transfer per unit length through the
interface $x=0$ is approximately equal to $3c_1 - 3c^+ = 3\Delta c$ (the factor
3 is the number of velocity directions in the LGA model that transfer particles
through the interface). This must be balanced by particle current caused by the
fluid flow, which can be approximated by $7c^+ \vf$ (the factor 7 is the number
of possible particle velocities). Thus, $3c_1 - 3c^+ \approx 7c^+\vf$, or
\begin{equation}
  \label{eq:Dc-Dv}
    \Delta c \approx  \frac{7c_1\vf}{3 + 7\vf}.
\end{equation}
Comparison of this formula with the actual data obtained in simulations is
shown in Figure (\ref{fig:Dc-Dv}). The agreement is fairly good.

\section{Discussion}
\label{sec:discussion}

Several features of the model deserve closer attention. First, the
model employs specular scatterers to mimic a porous medium as well as
to reduce hydrodynamic effects~\cite{kuntzJPD03}. However, the effect
of such scatterers on the diffusion along the ``x'' axis is very
peculiar: the ``x'' component of the momentum is changed only for
those particles, whose velocity vector is parallel to the ``x'' axis.
For this reason specular scatterers are widely used in LGA models to
implement slip-free walls parallel to the x axis. However, in the KL
model the scatterers are distributed randomly with a relatively small
concentration to form isolated, point-like islands. Such scatterers
are inefficient in breaking temporal correlations in the ``x''
component of momentum of particles that hit them. Thus, on the one
hand, randomly distributed specular scatterers have a very limited
impact on the diffusivity along the ``x'' axis. On the other hand,
the mean-free path, and hence the diffusion coefficient in the
original, scatterer-free FHP model diverge to infinity as the reduced
particle concentration goes to $0$ or $1$ \cite{kuntzJPD03}.
Combination of these two effects explains the fundamental property of
the KL model: extremely strong dependence of the diffusion
coefficient on concentration \cite{kuntzJPD03}.

Second, the present model is based on the FHP model, where particle-particle
collisions preserve mass and momentum to mimic fluid flow. Local conservation
of momentum leads to hydrodynamic flow from high-pressure (high-concentration)
to low-pressure (low-concentration) areas. Therefore, hydrodynamical flow in
any FHP-based model of porous media is inevitable and its effects should be
always carefully taken into account, especially when specular scatterers are
used to model a porous medium.

Third, assume that as soon as the flow velocity in the front becomes
sufficiently small, the particle flux $q$ associated with it obeys Darcy's law
\cite{BearBook}
\begin{equation}
  q = -\frac{\kappa}{\mu} \nabla P,
\end{equation} where $\kappa$ is the permeability of the medium,
$\mu$ denotes the fluid viscosity, and $\nabla P$ is the pressure
gradient. Taking into account that $q \propto v_x$ \cite{BearBook},
$\mu \propto D^{-1}$ (Stokes-Einstein relation), and $P \propto c$
\cite{DrozBook}, one arrives at \begin{equation}
\label{eq:vx_propto_c}
  v_x \propto D \nabla c.
\end{equation}
If $D$ is very large, even a very small concentration gradient can cause a
significant fluid flow.  Therefore, one can expect strong hydrodynamical
effects to appear in \emph{any} hydrodynamical model of porous medium with
diffusion coefficient $D(c)$ diverging to infinity at some value(s) of $c$.

% This observation sheds a
%new light on the cause of ``anomalous'' properties of the present model. While

%Third, strong hydrodynamical effects should appear in any hydrodynamical model
%of porous medium with diffusion coefficient diverging to infinity at some
%particle concentrations. This conjecture is based on an assumption that as soon
%as the flow velocity in the front becomes sufficiently small, the associated
%particle flux approximately follows Darcy's law
%\begin{equation}
%  q = -\frac{\kappa}{\mu} \nabla P,
%\end{equation}
%where $\kappa$ is the permeability of the medium, $\mu$ denotes the fluid
%viscosity, and $\nabla P$ is the pressure gradient.
%

Fourth, the value of $|x_0|$ in the similarity relation
(\ref{eq:scaling-moving}) turned out quite large. Similarly, the relaxation
time to the asymptotic, long-time regime in which the similarity holds,
$\tau_\mathrm{relax}$, is also large. These two effects may be related to the fact
that in
all FHP models $v_x$, which constitutes the left-hand side of relation
(\ref{eq:vx_propto_c}), is limited from above. However, the right-hand side of
(\ref{eq:vx_propto_c}) is a product of two quantities that at early stages of
the front propagation assume very large values in the front region. Thus, the
system can obey Darcy law only after a long relaxation time necessary to reduce
the concentration gradient $\nabla c$ to a value of order $1/D$. This explains
why in the previous studies the front dynamics was found ``anomalous'' only if
both $c_1$ was close to 1 and the specular scatterers were used
\cite{kuntz-rapidcomm,kuntzJPD03,kuntz-PRE05}. If any of these conditions was
not satisfied, the value of diffusion coefficient $D$ took on relatively small
values in the front region and the relaxation time was small enough to allow
the simulations to reach the asymptotic regime. Note also that large values of
$|x_0|$ and $\tau_\mathrm{relax}$ are in accord with several recent experiments
on sorptivity of building materials, where a short-time deviation from the
$t^{1/2}$ behavior was found (see \cite{Lockington} and references therein).

%Fifth, the negative sign of $x_0$ is a consequence of the
%fact that the actual front velocity at
%early times is smaller than its value extrapolated from the long-time, ``Darcy''
%asymptotics to early times (see Fig.~\ref{fig:v-t}).

Fifth, the actual front velocity at
early times turned out smaller than its value extrapolated from the long-time, ``Darcy''
asymptotics  (see Fig.~\ref{fig:v-t}). As a consequence,
$x_0$ is negative and the front dynamics at early times appears ``superdiffusive''
($\alpha > 1/2$). In systems where the early-time velocity
is larger than the short-time extrapolation of the Darcy asymptotics,
$x_0$ would be positive and the early-time front dynamics would appear
``subdiffusive'' ($\alpha < 1/2$).

Finally, Figure \ref{fig:v-t} suggests that at large times the mean
front velocity, $\vf$, becomes proportional to $1/\sqrt{t}$, which
yields a characteristic length scale $\int_{\tau=0}^t
1/\sqrt{\tau}\,\mathrm{d}\tau \propto \sqrt{t}$. This implies that
the characteristic lengths associated to both diffusion and fluid
flow asymptotically scale with time in the same way, as $\sqrt t$.
Consequently, the KL model is an example of a system where
measurements based only  on analysis of asymptotic concentration
profiles cannot differentiate between diffusive and hydrodynamic
effects.

\section{Conclusions}
\label{sec:conclusions}

We have shown that the unusual dynamics of the concentration front in
the KL model of a porous medium can be fully explained as a combined
effect of hydrodynamic flow and normal diffusion. Our argumentation
is not only simpler than that proposed by K\"{u}ntz and Lavall\'{e}e
\cite{kuntz-rapidcomm,kuntzJPD03,kuntz-PRE05}, who used the concept
of anomalous diffusion, but also allows to explain more physical
properties of the model, e.g.\ a concentration drop at the boundary
$x=0$.

From our analysis the following picture of the concentration front dynamics in
the model emerges. A concentration gradient across the front leads to a
pressure gradient which, in turn, forces the FHP fluid to flow. This flow is
attenuated by scatterers that mimic a porous medium, so that hydrodynamical
effects in most cases are negligible, especially at large times. However,
randomly distributed specular scatterers used in the KL model are very
inefficient at reduced concentrations close to 1. In this particular case the
viscosity goes to 0, and both diffusivity and mobility along the ``x'' axis
tend to infinity.
Under these conditions,  hydrodynamical flow significantly affects the front
dynamics at all times. In particular, it renders the relaxation time to the
asymptotic regime, where Darcy law holds, very large, and significantly changes
the position of the Boltzmann-Matano interface $x_0$. The front dynamics is
controlled by both diffusion and fluid flow, and at very large times the
characteristic length scales of both processes become proportional to
$\sqrt{t}$. Consequently, at very large times the concentration profile can be
expressed as a function of a single variable $x/\sqrt{t}$, which could easily
be misinterpreted that the front dynamics is governed only by diffusion.

While the KL model cannot be regarded as a model of anomalous
diffusion, it remains an interesting model of fluid flow in a porous
medium. Its main advantage is ability to reproduce some non-trivial
properties of concentration profiles found in several recent
experiments. Similarity between the model an experimental results
suggests that the anomalous front behavior observed e.g. in
experiments on building materials, may have nothing to do with
anomalous diffusion, but is caused by some hydrodynamic effects
that accompany normal diffusion.

%\section*{Acknowledgments}

\section*{References}

%\bibliography{porowate}

\begin{thebibliography}{10}

\bibitem{ghandy}
A.~El-Ghany~El Abd and J.J. Milczarek.
\newblock Neutron radiography study of water absorption in porous building
  materials: anomalous diffusion analysis.
\newblock {\em J. Phys. D: Appl. Phys.}, 37:2305, 2004.

\bibitem{pel}
L.~Pel, K.~Kopinga, G.~Bertram, and G.~Lang.
\newblock Water absorption in in a fired-clay brick observed by \uppercase{NMR}
  scanning.
\newblock {\em J. Phys. D: Appl. Phys.}, 34:675, 1995.

\bibitem{kuntz-JPD01}
M.~K\"{u}ntz and P.~Lavall\'{e}e.
\newblock Experimental evidence and theoretical analysis of anomalous diffusion
  during water infiltration in porous building materials.
\newblock {\em J. Phys. D: Appl. Phys.}, 34:2547, 2001.

\bibitem{Lockington}
D.A. Lockington and J.-Y. Parlange.
\newblock Anomalous water absorption in porous materials.
\newblock {\em J. Phys. D: Appl. Phys.}, 36:760, 2003.

\bibitem{Carey95}
A.E. Carey, S.W. Wheatcraft, R.J. Glass, and J.P. O'Rourke.
\newblock Non-\uppercase{F}ickian ionic diffusion across high-concentration
  gradients.
\newblock {\em Water Resour. Res.}, 31:2213, 1995.

\bibitem{kuntz-rapidcomm}
M.~K\"{u}ntz and P.~Lavall\'{e}e.
\newblock Anomalous diffusion is the rule in concentration-dependent diffusion
  processes.
\newblock {\em J. Phys. D: Appl. Phys.}, 37:L5--L8, 2004.

\bibitem{Rehage70}
G.~Rehage, O.~Ernst, and J.~Fuhrmann.
\newblock Fickian and non-fickian diffusion in high polymer systems.
\newblock {\em Discuss. Faraday Soc.}, 49:208--221, 1970.

\bibitem{Fowlkes}
J.D.~Fowlkes \emph{et al}.
\newblock Molecular transport in a crowded volume created from vertically
  aligned carbon nanofibres: a fluorescence recovery after photobleaching
  study.
\newblock {\em Nanotechnology}, 17:5659, 2006.

\bibitem{Naumovec}
A.T. Lobutets, A.G. Naumovets, and Yu.~S. Vedula.
\newblock Diffusion of adsorbed particles on surfaces with channeled atomic
  corrugation.
\newblock In A.~P\c{e}kalski and K.~Sznajd-Weron, editors, {\em Anomalous
  Diffusion: From Basics to Applications}, volume~59 of {\em Lecture Notes in
  Physics}, pages 340--348. Springer, 1999.

\bibitem{Gomer}
R.~Gomer.
\newblock Diffusion of adsorbates on metal surfaces.
\newblock {\em Rep. Prog. Phys.}, 53:917, 1990.

\bibitem{HavlinBook}
D.\ ben Avraham and S.\ Havlin.
\newblock {\em Diffusion and Reactions in Fractals and Disordered Systems}.
\newblock Cambridge Univ.\ Press, Cambridge, 2000.

\bibitem{Metzler-Klafter}
R.~Metzler and J.~Klafter.
\newblock The random walk's guide to anomalous diffusion: a fractional dynamics
  approach.
\newblock {\em Phys. Rep.}, 339(1):1--77, 12 2000.

\bibitem{kuntzJPD03}
M.~K\"{u}ntz and P.~Lavall\'{e}e.
\newblock Anomalous spreading of a density front from an infinite continuous
  source in a concentration-dependent lattice gas automaton diffusion model.
\newblock {\em J. Phys. D: Appl. Phys.}, 36:1135, 2003.

\bibitem{kuntz-PRE05}
M.~K\"{u}ntz and P.~Lavall\'{e}e.
\newblock Numerical investigation of the spreading-receding cycle in a
  concentration-dependent lattice gas automaton diffusion model.
\newblock {\em Phys. Rev. E}, 71:066703, 2005.

\bibitem{kuntzTPM01}
M.~K\"{u}ntz, J.G.M. van Mier, and P.~Lavall\'{e}e.
\newblock A lattice gas automaton simulation of the non-linear diffusion
  equation: a model for moisture flow in unsaturated porous media.
\newblock {\em Transport in Porous Media}, 43:289, 2001.

\bibitem{frish-lga}
U.~Frish, B.~Hasslacher, and Y.~Pomeau.
\newblock Lattice-gas automata for the \mbox{N}avier-\mbox{S}tokes equation.
\newblock {\em Phys. Rev. Lett.}, 56:1505, 1986.

\bibitem{CrankBook}
J.\ Crank.
\newblock {\em The mathematics of diffusion}.
\newblock Oxford Univ. Press, Oxford, 1956.

\bibitem{BearBook}
J.~Bear.
\newblock {\em Dynamics of Fluids in Porous Media}.
\newblock Courier Dover Publications, Dover, 1989.

\bibitem{DrozBook}
B.~Chopard and M.~Droz.
\newblock {\em Cellular Automata Modeling of Physical Systems}.
\newblock Cambridge Univ. Press, Cambridge, 1998.

\end{thebibliography}

\end{document}